\documentclass[conference]{IEEEtran}
\IEEEoverridecommandlockouts
\usepackage{cite}
\usepackage{amsmath,amssymb,amsfonts}
\usepackage{algorithmic}
\usepackage{graphicx}
\usepackage{textcomp}
\usepackage{xcolor}
\usepackage{color,colortbl}
\usepackage{algorithm}
\usepackage[vlined, ruled, algo2e]{algorithm2e}
\usepackage{subcaption}
\DeclareCaptionSubType*[Alph]{table}
\DeclareCaptionLabelFormat{mystyle}{Table~\bothIfFirst{#1}{ ̃}#2}
\def\BibTeX{{\rm B\kern-.05em{\sc i\kern-.025em b}\kern-.08em
    T\kern-.1667em\lower.7ex\hbox{E}\kern-.125emX}}
\begin{document}

\title{Identifying Indoor Points of Interest via Mobile Crowdsensing: An Experimental Study}
\makeatletter
\newcommand{\linebreakand}{%
\end{@IEEEauthorhalign}
\hfill\mbox{}\par
\mbox{}\hfill\begin{@IEEEauthorhalign}
}
\makeatother
\author{\IEEEauthorblockN{Sumudu Hasala Marakkalage}
\IEEEauthorblockA{\textit{Engineering Product Development Pillar} \\
\textit{Singapore University of Technology and Design}\\
Singapore \\
marakkalage@mymail.sutd.edu.sg}
\and
\IEEEauthorblockN{Ran Liu}
\IEEEauthorblockA{\textit{Engineering Product Development Pillar} \\
\textit{Singapore University of Technology and Design}\\
Singapore \\
ran\_liu@sutd.edu.sg}
\linebreakand
\IEEEauthorblockN{Sanjana Kadaba Viswanath}
\IEEEauthorblockA{\textit{Engineering Product Development Pillar} \\
\textit{Singapore University of Technology and Design}\\
Singapore \\
kadabasanju@gmail.com}
\and
\IEEEauthorblockN{Chau Yuen}
\IEEEauthorblockA{\textit{Engineering Product Development Pillar} \\
\textit{Singapore University of Technology and Design}\\
Singapore \\
yuenchau@sutd.edu.sg}
}

\maketitle

\begin{abstract}
This paper presents a mobile crowdsensing approach to identify the indoor points of interest (POI) by exploiting Wi-Fi similarity measurements. Since indoor environments are lacking the GPS positioning accuracy when compared to outdoors, we rely on widely available Wi-Fi access points (AP) in contemporary urban indoor environments, to accurately identify user POI. We propose a smartphone application based system architecture to scan the surrounding Wi-Fi AP and measure the cosine similarity of received signal strengths (RSS), and demonstrate through the experimental results that it is possible to identify the distinct POI of users, and the common POI among users of a given indoor environment. 
\end{abstract}


\section{Introduction}
In recent years, mobile crowdsensing has gained a significant interest due to the ubiquitous nature of smartphones, their in-built sensing capabilities, and the fact that they are carried by humans. Therefore, mobile crowdsensing applications have been envisioned in diverse domains such as transportation \cite{farkas2015crowdsending}, healthcare \cite{leonardi2014secondnose}, and social networking \cite{hu2014multidimensional}. Proper crowd participation enables fine-grained spatio-temporal monitoring of a particular phenomenon by harnessing diverse information through smartphone-based applications \cite{hoteit2014estimating},\cite{kang2013exploring}. Especially in human mobility tracking, knowing both outdoor and indoor mobility patterns gives important insights into the user behavior analysis \cite{li2008mining}, \cite{lou2009map}. Identifying the places where a particular user visits (points of interest) is important when studying mobility patterns and providing context-aware services. Human trajectory analysis to learn the motion patterns and to detect anomalies is done in \cite{suzuki2007learning}. Previous work has done to detect the surrounding environment type (e.g. Indoor/Outdoor) using smartphone based sensors \cite{zhou2012iodetector}, \cite{7917558}. Understanding the lifestyle of older adults is conducted in \cite{marakkalage2018understanding} through smartphone application data. It mainly focuses on extracting regions of interest (ROI) and points of interest (POI), based on sensor fusion. However, in large indoor environments such as shopping malls or offices, those work lack the indoor POI granularity. Indoor location tracking is challenging due to the difficulty in obtaining fine-grained location information in indoor environments. Since the contemporary urban environments are equipped with dense Wi-Fi access points (AP), those AP can be exploited to identify the indoor POI. Previous research has been conducted using Wi-Fi AP information to create indoor floorplans \cite{alzantot2012crowdinside} and indoor localization \cite{zhu2014spatio}, \cite{liu_crowdsensing_2019}, \cite{liu_ieee_sensors2017}. However, those require higher sampling rate of data collection and it leads to high power consumption in mobile devices, which is a prime challenge in mobile crowdsensing \cite{ganti2011mobile}, \cite{lau2019survey}. Also, it requires an extensive labour cost when creating fingerprint maps of indoor environments. 

We focus on the typical user behaviour in an indoor environment such as shopping mall or office/school, where users have a set of POI they frequently visit. Therefore, the main objective of this paper is to identify the distinct user POI in indoor environments, and to identify how much time users spend in each POI. Our focus is to verify the effectiveness of the proposed system, based on the data collected from volunteer users along with the ground truth of the POI they visited. The contributions of this paper are listed as follows.
\begin{itemize}
	\item Design and development of an efficient mobile crowdsensing system architcture to collect the surrounding Wi-Fi AP information using a smartphone application and analyse those information to extract indoor POI. 
	\item Identifying the distinct POI of a particular user in a given indoor environment, and identifying the revisited POI by the same user.
	\item Identifying the common POI among multiple users.
\end{itemize}

The subsequent sections of this paper are organized as follows. In Section \ref{sec_2}, the overview of the proposed system is presented. In Section \ref{sec_3}, the POI extraction process is presented along with the technical details. In Section \ref{sec_4}, the results of an experimental study are presented. Section \ref{sec_4} concludes the paper with a discussion and future work.

\section{System Overview}
\label{sec_2}

The proposed system consists of a front-end smartphone application, which collects surrounding Wi-Fi AP information and sends them to a back-end server application, which analyses raw Wi-Fi data to identify the indoor POI and the time duration, in which a particular user stayed at each POI. An overview of the proposed system is shown in Figure \ref{fig:overview}.

\begin{figure}[!htb] 
	\centering
	\includegraphics[width=0.5\textwidth]{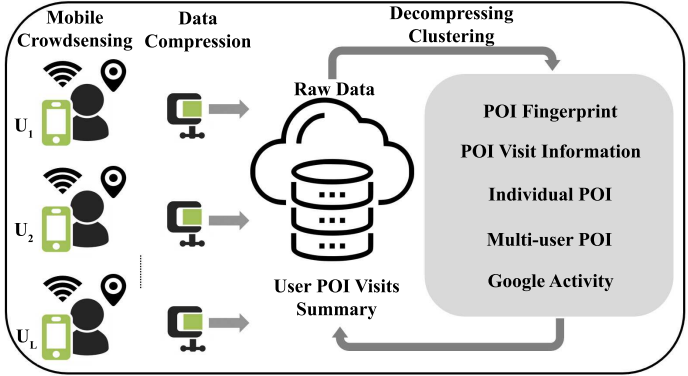} 
	\caption{Overview of the proposed system} 
	\label{fig:overview} 
\end{figure}

\subsection{Front-end}
The front-end is an Android smartphone application, which collects the smartphone location data (GPS coordinates) and the surrounding Wi-Fi AP information (MAC address and corresponding RSS value) by a background service.

\subsubsection{Wi-Fi Scanning}

Scanning of the surrounding Wi-Fi AP information is done every $5$ minutes, since excessive Wi-Fi scanning impacts heavily on smartphone battery consumption, according to the Android API \cite{wifiscan}. Let MAC address of the AP be $m$, and the received signal strength (RSS) of the AP be $r$ in $dBm$. The scan result ($s$) is a list of surrounding AP MAC addresses and their corresponding (RSS) as shown in the Equation \ref{eq_scanResult}, where $n$ is the number of AP obtained in a given scan result.

\begin{equation}
\label{eq_scanResult}
s = {\{m_1,r_1\},\{m_2,r_2\},...,\{m_n,r_n\}}
\end{equation}

Each scan result is stored with the corresponding timestamp ($t$) of the Wi-Fi scan. 
The list of scan results ($S$) is denoted as shown in Equation \ref{eq_listOfScanResults}, where $m$ is the number of scan results in the list. 

\begin{equation}
\label{eq_listOfScanResults}
S = {\{s_1,t_1\},\{s_2,t_2\},...,\{s_m,t_m\}}
\end{equation}

The obtained list of scan results is stored in the device until it is uploaded to cloud server for further processing. Once the data is uploaded successfully, the list is cleared to preserve smartphone storage space.     

\subsubsection{Data Compression}
Transmission of raw data into the cloud server adds an extensive network cost. To overcome this problem, we compress the raw data using \textit{gzip} \cite{gzip}, which uses Huffman coding technique. As shown in Table \ref{tab_compression}, we notice that a $6$ hour duration of data becomes significantly smaller in size when compressed. Therefore, we upload compressed raw data to server every $6$ hours, provided that the device is connected to a Wi-Fi network. If there is no Wi-Fi connection, the smartphone application will wait until it connects to one.

\begin{table}[!htb]
\centering
\caption{Data size before and after compression}
\label{tab_compression}	
	\begin{tabular}{|c|c|c|}
		\hline
		\textbf{\begin{tabular}[c]{@{}c@{}}Duration \\ (Hours)\end{tabular}} & \textbf{\begin{tabular}[c]{@{}c@{}}Uncompressed\\ (Bytes)\end{tabular}} & \textbf{\begin{tabular}[c]{@{}c@{}}Compressed\\ (Bytes)\end{tabular}} \\ \hline
		0.5                                                                  & 20,701                                                                   & 656                                                                   \\ \hline
		1                                                                    & 41,401                                                                   & 791                                                                   \\ \hline
		3                                                                    & 172,501                                                                  & 1,562                                                                  \\ \hline
		6                                                                    & 345,001                                                                  & 2,565                                                                  \\ \hline
	\end{tabular}
\end{table}

\subsection{Back-end}
When the compressed raw data is received at back-end server, it is further processed to identify the indoor POI.
A database model as shown in Figure \ref{fig:datamodel} is used to store the raw data and processed data. It consists of two major components, namely raw database and summary database. 
\begin{figure}[!htb] 
	\centering
	\includegraphics[width=0.5\textwidth]{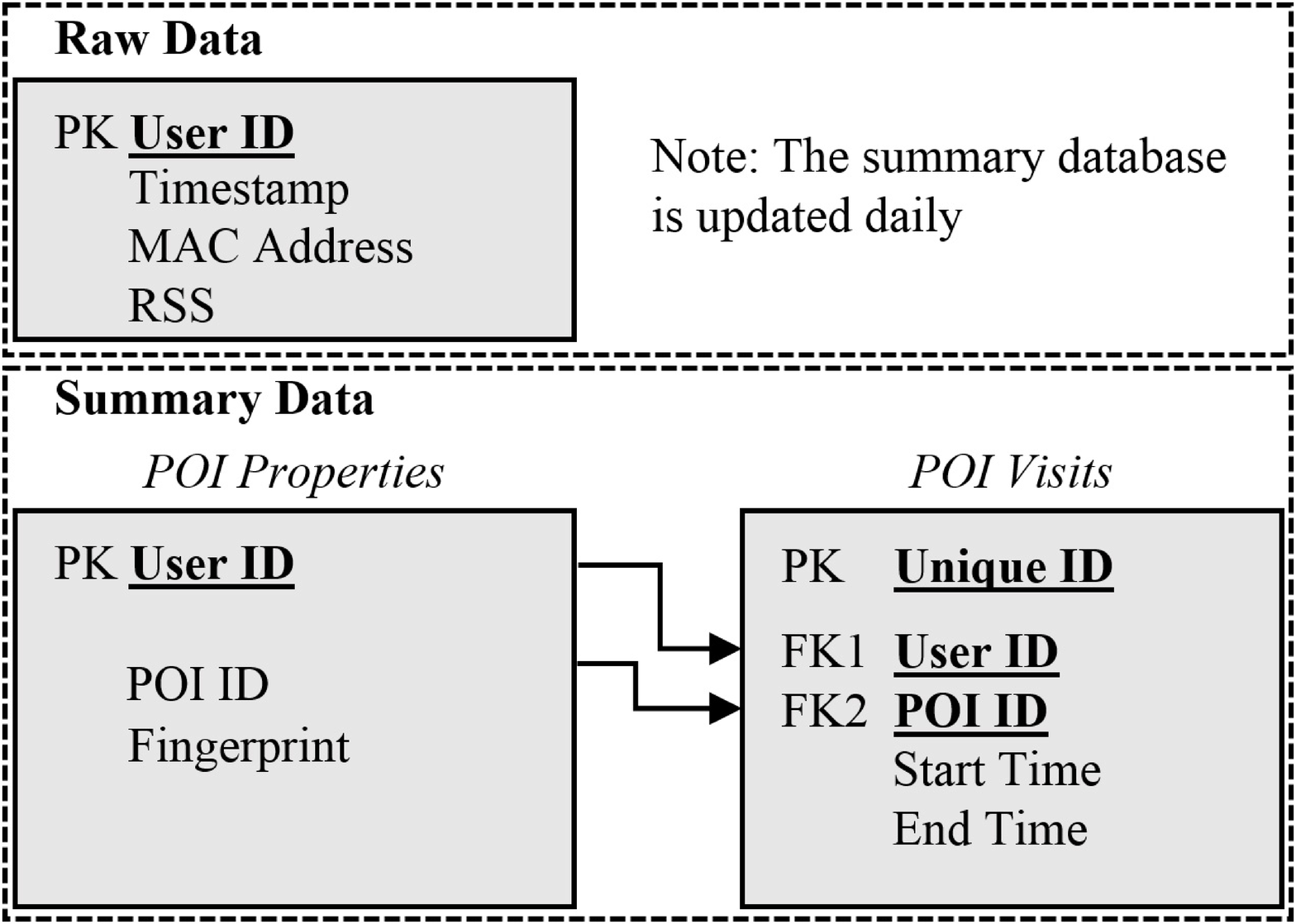} 
	\caption{Database model in the back-end server, where PK is the primary key, and FK is the foreign key} 
	\label{fig:datamodel} 
\end{figure}
\section{Indoor POI Extraction}
\label{sec_3}
Indoor POI extraction is done using Wi-Fi fingerprint clustering and matching their cosine similarity scores. The following subsections present the clustering technique and the similarity measures used in user POI extraction.
\subsection{Clustering Technique}
According to the work done in \cite{7917558}, different clustering algorithms are experimentally evaluated, and DBSCAN \cite{ester1996density} has been chosen as the preferred clustering technique due to its capabilities in forming arbitrary shape clusters. In this paper we propose a modified DBSCAN technique to cluster the raw Wi-Fi RSS measurements for a particular device in a given environment. We used cosine similarity measure between RSS values as the distance metric of DBSCAN algorithm. Based on typical user behaviour, we form a POI if a user stays at least $20$ minutes in one place. Therefore, we select DBSCAN parameters such as minimum points to form a cluster ($minPts$) as $4$ (based on $5$ minute scan interval), and the similarity threshold ($\epsilon$) as $0.5$ (based on experimental evaluation as shown in the Table \ref{tab:simThreshold}). Algorithm \ref{algo:POI} shows the clustering procedure for a given set of input raw Wi-Fi data ($S$), the similarity threshold $\epsilon$, and the minimum points required to form a cluster $minPts$. The output list of cluster points is $P$.

\begin{algorithm}[]
	\SetAlgoLined
	\KwIn{similarity threshold ($\epsilon$), $minPts$, Wi-Fi list ($S$)} 
	\KwOut{Cluster point list ($P$)}
	Visited points ($ V_p$), index $(z_1)$, $P = 0$ \\
	\While{size of $S$ $\ge$ $z_1$}{
		$\alpha =S[z_i]$\\
		\eIf{$\alpha \not\subset V_p$}{
			add $\alpha$ to $V_p$\\
			$N$ = get neighbours of $\alpha$\\
			\If{size of $N \geq minPts$}
			{$z_2=0$\\
				\While{size of $N \geq z_2$}
				{$\beta = N[z_2]$\\
					\eIf{$\beta \not\subset V_p$}
					{add $\beta$ to $V_p$\\
						$Q$ = get neighbours of $\beta$\\
						\If{size of $Q \geq minPts$}{merge $Q$ with $N$ }{}
						}
					{$z_2 = z_2+1$}
					}
					add $N$ to $P$
				
				}
			{} 
		}{
		$z_1 = z_1+1$\;
	}
}
\caption{POI extraction from raw Wi-Fi data}
\label{algo:POI}
\end{algorithm}

The process of finding the neighbour points is shown in Algorithm \ref{algo:neighbours}, where inputs are $\alpha$ and $S$, and the output is $N$, which are mentioned in Algorithm \ref{algo:POI}.  A detailed explanation on the similarity metric is presented in Section \ref{sim_metric} 

\begin{algorithm}[!htb]
	\SetAlgoLined
	\KwIn{Scan result ($\alpha$),Wi-Fi list ($S$)}
	\KwOut{Neighbour points ($N$)}
	$N = 0$\\
	\For{every index i in S}
	{$D = $ calculate similarity($\alpha,S[i]$)\\
		\If{$D \geq \epsilon$}{add $S[i]$ to $N$}
		}
	
\caption{The process of finding neighbour points}
\label{algo:neighbours}
\end{algorithm}

Once the clustering process is completed, a fingerprint for each distinct POI is generated along with a unique POI ID and saved in the ``POI Properties" table in summary database. The POI fingerprint ($F$) is denoted as shown in the Equation \ref{eq_fingerprint}, where $M$ is the MAC address, $R$ is the corresponding average RSS value in $dBm$, and $p$ is the number of distinct MAC addresses detected at that particular POI.

\begin{equation}
\label{eq_fingerprint}
F = {\{M_1,R_1\},\{M_2,R_2\},...,\{M_p,R_p\}}
\end{equation}

\subsection{Similarity Metric}
\label{sim_metric}
We used cosine similarity as the distance metric in DBSCAN algorithm. The cosine similarity between two Wi-Fi fingerprints $F_1$ and $F_2$ is calculated as follows.
\begin{equation}
\label{eq_f1}
F_1 = {\{M^1_1,R^1_1\},\{M^1_2,R^1_2\},...,\{M^1_u,R^1_u\}}
\end{equation}
\begin{equation}
\label{eq_f2}
F_2 = {\{M^2_1,R^2_1\},\{M^2_2,R^2_2\},...,\{M^2_v,R^2_v\}}
\end{equation}
where $u$ and $v$ are the number of distinct MAC addresses in $F_1$ and $F_2$ respectively. The dot product of the RSS values of common MAC addresses in the two fingerprints ($Y$) is calculated as shown in Equation \ref{eq_dotcommon}, where $w$ is the number of common MAC addresses.
\begin{equation}
\label{eq_dotcommon}
Y = \sum_{i=1}^{w} [R^1_i \cdot R^2_i]
\end{equation}
The dot products of each RSS value of $F_1$ and $F_2$ are shown in the Equations \ref{eq_d1} and \ref{eq_d2} respectively.
\begin{equation}
\label{eq_d1}
d_1 = \sum_{j=1}^{u} [R^1_j \cdot R^1_j]
\end{equation}
\begin{equation}
\label{eq_d2}
d_2 = \sum_{k=1}^{v} [R^2_k \cdot R^2_k]
\end{equation}
Finally, the cosine similarity score ($C$) between the two fingerprints is calculated as shown in the Equation \ref{eq_cosine}.
\begin{equation}
\label{eq_cosine}
C = Y/ (\sqrt{d_1} \times \sqrt{d_2})  \text{ ; where } 0 \leq C \leq 1
\end{equation}

\subsection{Community Detection (Common POI Among Users)}
Once we obtain the indoor POI for individual users, knowing the common POI among users is useful for user mobility analysis. We exploit Louvain method for community detection \cite{blondel2008fast} to obtain insights about popular POI among users. Let the number of POI in a given indoor environment is $h$, the number of pair-wise cosine similarities (I) is calculated as shown in the Equation \ref{eq_pairwise}. 
\begin{equation}
\label{eq_pairwise}
I = \frac{h!}{r!(h-r)!}\quad \text{; where } r=2
\end{equation}
$I$ is given as input to Louvain algorithm, which tries to find the best partitioning among the POI (nodes) comparing pair-wise similarity (edges), and gives a modularity score. The community detection resuls from our experimental study is presented in Section \ref{subsec:commonPOI}.  
\section{Experimental Study}
\label{sec_4}

We collected crownsensing data using the aforementioned smartphone application, from a group of people who use different models of smartphones along with the ground truth of the indoor POI they visited. The following subsections present the experimental results obtained in different scenarios of indoor POI extraction. 
\subsection{Impact of Different Similarity Threshold Values}
\label{sec:similarity_evaluation}
We have experimentally evaluated the impact of different cosine similarity threshold values for indoor POI extraction. As shown in the Table \ref{tab:simA} and Table \ref{tab:simB}, the performance of two different similarity threshold values are evaluated along with the ground truth POI labels. When the similarity threshold is $0.5$, the indoor POI extraction matches with the ground truth. An additional POI is identified (POI ID $5$) when the similarity threshold is $0.6$ as highlighted in the Table \ref{tab:simB}. This shows that when the similarity threshold gets higher, even the changes in environment (e.g. movement of people) could impact on the RSS values scanned by the smartphone application.  
\begin{table}[]
	\centering
	\caption{Impact of different cosine similarity threshold values when extracting indoor POI}
	\label{tab:simThreshold}
	\begin{subtable}{0.5\textwidth}
		\centering
		\caption{Similarity threshold $(\epsilon) = 0.5$}
		\label{tab:simA}
\begin{tabular}{|c|l|c|c|}
	\hline
	\textbf{\begin{tabular}[c]{@{}c@{}}POI \\ ID\end{tabular}} & \multicolumn{1}{c|}{\textbf{\begin{tabular}[c]{@{}c@{}}Ground Truth \\ Label\end{tabular}}} & \textbf{\begin{tabular}[c]{@{}c@{}}Start Time\\ (HH:mm)\end{tabular}} & \textbf{\begin{tabular}[c]{@{}c@{}}End Time\\ (HH:mm)\end{tabular}} \\ \hline
	1                                                          & Home                                                                                        & 00:00                                                                 & 09:23                                                               \\ \hline
	2                                                          & Office                                                                                      & 09:59                                                                 & 11:34                                                               \\ \hline
	3                                                          & Meeting room                                                                                & 11:58                                                                 & 14:57                                                               \\ \hline
	4                                                          & Canteen                                                                                     & 15:29                                                                 & 16:39                                                               \\ \hline
	2                                                          & Office                                                                                      & 16:44                                                                 & 17:09                                                               \\ \hline
	1                                                          & Home                                                                                        & 17:49                                                                 & 23:53                                                               \\ \hline
\end{tabular}
	\vspace{0.2cm}
	\end{subtable}
	
	\begin{subtable}{0.5\textwidth}
		\centering
		\caption{Similarity threshold $(\epsilon) = 0.6$}
		\label{tab:simB}
\begin{tabular}{|c|l|c|c|}
	\hline
	\textbf{\begin{tabular}[c]{@{}c@{}}POI \\ ID\end{tabular}} & \multicolumn{1}{c|}{\textbf{\begin{tabular}[c]{@{}c@{}}Ground Truth \\ Label\end{tabular}}} & \textbf{\begin{tabular}[c]{@{}c@{}}Start Time\\ (HH:mm)\end{tabular}} & \textbf{\begin{tabular}[c]{@{}c@{}}End Time\\ (HH:mm)\end{tabular}} \\ \hline
	1                                                          & Home                                                                                        & 00:00                                                                 & 09:23                                                               \\ \hline
	2                                                          & Office                                                                                      & 09:59                                                                 & 11:34                                                               \\ \hline
	3                                                          & Meeting room                                                                                & 11:58                                                                 & 14:57                                                               \\ \hline
	4                                                          & Canteen                                                                                     & 15:29                                                                 & 16:39                                                               \\ \hline
	2                                                          & Office                                                                                      & 16:44                                                                 & 17:09                                                               \\ \hline
	1                                                          & Home                                                                                        & 17:49                                                                 & 18:54                                                               \\ \hline
	\rowcolor[HTML]{FD6864} 
	5                                                          & Home                                                                                        & 18:59                                                                 & 20:39                                                               \\ \hline
	1                                                          & Home                                                                                        & 20:48                                                                 & 23:53                                                               \\ \hline
\end{tabular}
	\end{subtable}%
\end{table}

\subsection{Individual POI Matching}

For individual user POI matching, we conducted an experiment to identify the POIs when a particular user revisits them. We compared the obtained results along with the ground truth. Table \ref{tab_individualiuser} shows the individual user POI matching for user B in a particular day's POI visits. According to Table \ref{tab_individualiuser}, the system is capable of identifying when the user revisits (highlighted in same colour) POI ID $1$, $9$, and $12$. Those POI IDs are repeated because the system found that the cosine similarity score is higher than the designated $0.5$ threshold.

\begin{table}[]
\centering
\caption{Individual user POI matching scenario}
\label{tab_individualiuser}
\begin{tabular}{|l|l|c|c|}
	\hline
	\multicolumn{1}{|c|}{\textbf{\begin{tabular}[c]{@{}c@{}}POI \\ ID\end{tabular}}} & \multicolumn{1}{c|}{\textbf{\begin{tabular}[c]{@{}c@{}}Ground Truth \\ Label\end{tabular}}} & \textbf{\begin{tabular}[c]{@{}c@{}}Start Time\\ (HH:mm)\end{tabular}} & \textbf{\begin{tabular}[c]{@{}c@{}}End Time\\ (HH:mm)\end{tabular}} \\ \hline
	\rowcolor[HTML]{FFCCC9} 
	01                                                                             & Home                                                                                        & 00:00                                                                 & 08:45                                                               \\ \hline
	\rowcolor[HTML]{96FFFB} 
	12                                                                             & Canteen                                                                                     & 09:25                                                                 & 09:45                                                               \\ \hline
	\rowcolor[HTML]{CBCEFB} 
	09                                                                             & Office                                                                                      & 09:50                                                                 & 10:25                                                               \\ \hline
	60                                                                             & Corridor outside office                                                                     & 10:30                                                                 & 10:50                                                               \\ \hline
	\rowcolor[HTML]{CBCEFB} 
	09                                                                             & Office                                                                                      & 10:55                                                                 & 11:35                                                               \\ \hline
	\rowcolor[HTML]{96FFFB} 
	12                                                                             & Canteen                                                                                     & 11:40                                                                 & 12:05                                                               \\ \hline
	\rowcolor[HTML]{CBCEFB} 
	09                                                                             & Office                                                                                      & 12:10                                                                 & 17:50                                                               \\ \hline
	\rowcolor[HTML]{FFCCC9} 
	01                                                                             & Home                                                                                        & 18:00                                                                 & 23:59                                                               \\ \hline
\end{tabular}
\end{table}

\subsection{Common POI Among Users}
\label{subsec:commonPOI}

In the common POI matching scenario, we identify the popular POI, where multiple users prefer to visit. We choose a shopping mall, where $11$ users from the study visit for shopping/dining purposes. Users and the smartphone models used in the study are shown in the Table \ref{tab_smartphones}. 
A total of $41$ records of indoor POI at the shopping mall were detected. According to Equation \ref{eq_pairwise}, $820$ pair-wise similairites are fed to Louvain algorithm to build the community. We evaluated different similarity thresholds as shown in Table \ref{tab_modularity} to understand the best partition for communities, based on the Louvain modularity. Since, different POI has different sizes in space (e.g. Food court is larger when compared to a clothing shop), our objective is to identify even the smallest POI in the shopping mall visited by users. Therefore, for POI identication we used $0.5$ threshold in community detection. We notice that a large area like food court is divided into multiple POI. Figure \ref{fig:shopping_mall} shows the common POI among the $11$ users in a shopping mall environment.

\begin{table}[!htb]
	\centering
	\caption{Users and their smartphone models}
	\label{tab_smartphones}
	\begin{tabular}{|c|l|}
		\hline
		\textbf{User} & \multicolumn{1}{c|}{\textbf{Device Model}} \\ \hline
		A             & OnePlus 3                                  \\ \hline
		B,F             & Samsung Galaxy S8                                \\ \hline
		C             & Sony Xperia Z3                                \\ \hline
		D             & Google Pixel 2                                \\ \hline
		E             & Huawei Nova 2i                             \\ \hline
		G             & Xiaomi Max 2                             \\ \hline		
		H             & Oppo F5                          \\ \hline		
		I             & Oppo R11                          \\ \hline		
		J             & Xiaomi Mix 2                         \\ \hline		
		K             & Samsung Galaxy A8                         \\ \hline		
	\end{tabular}
\end{table}
\begin{table}[htb!]
	\centering
	\caption{Similarity threshold and Louvain modularity}
	\label{tab_modularity}
	\begin{tabular}{|c|c|}
		\hline
		\textbf{Threshold} & \textbf{Louvain Modularity} \\ \hline
		0.2                & 0.625                       \\ \hline
		0.3                & 0.803                       \\ \hline
		0.4                & 0.766                       \\ \hline
		0.5                & 0.692                       \\ \hline
	\end{tabular}
\end{table}

\begin{figure*}
	\centering
	\begin{subfigure}{0.32\textwidth}
		\includegraphics[width=\linewidth]{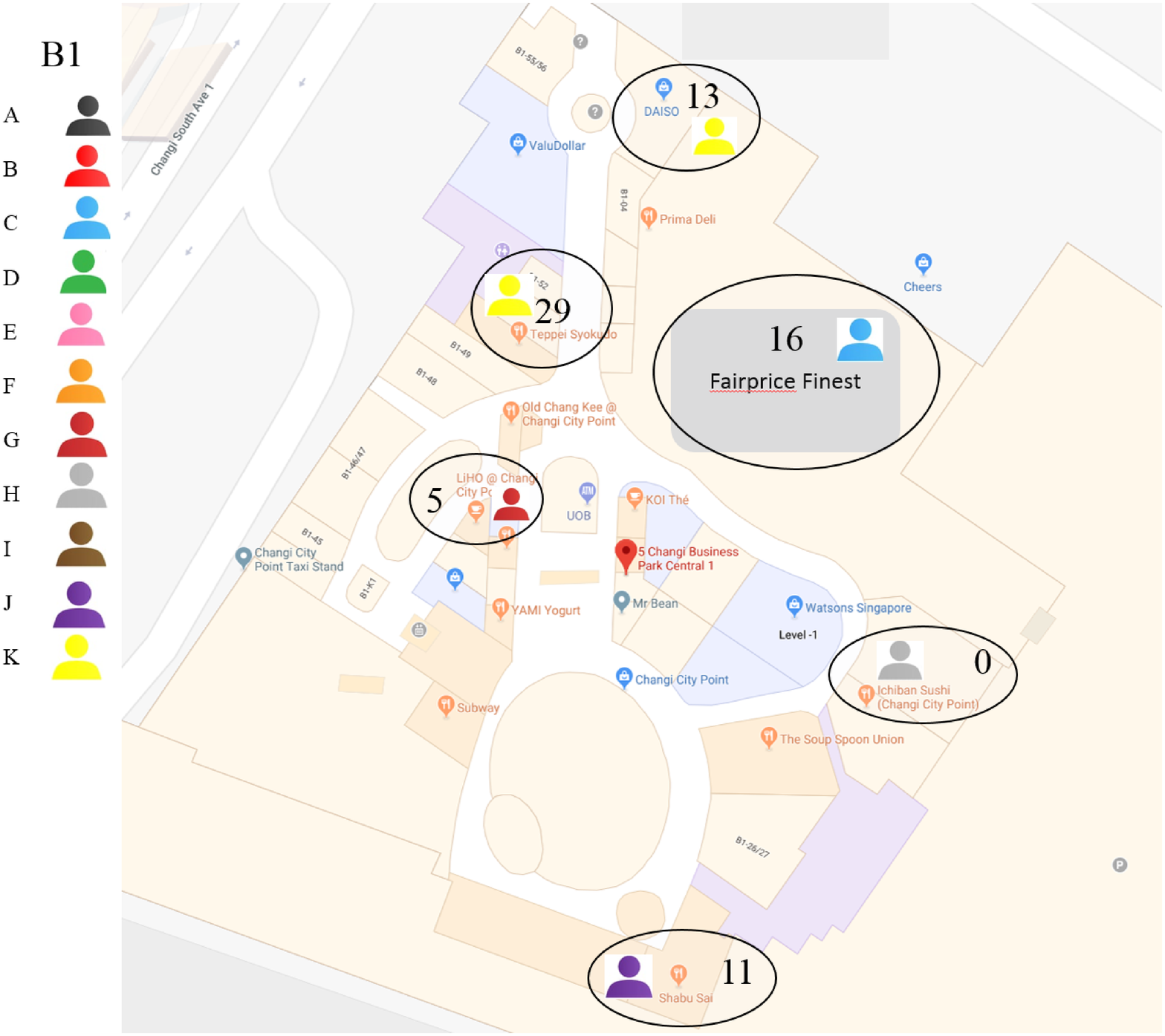}
		\caption{Basement 1} \label{fig:1a}
	\end{subfigure}
	\hspace*{\fill} 
	\begin{subfigure}{0.32\textwidth}
		\includegraphics[width=\linewidth]{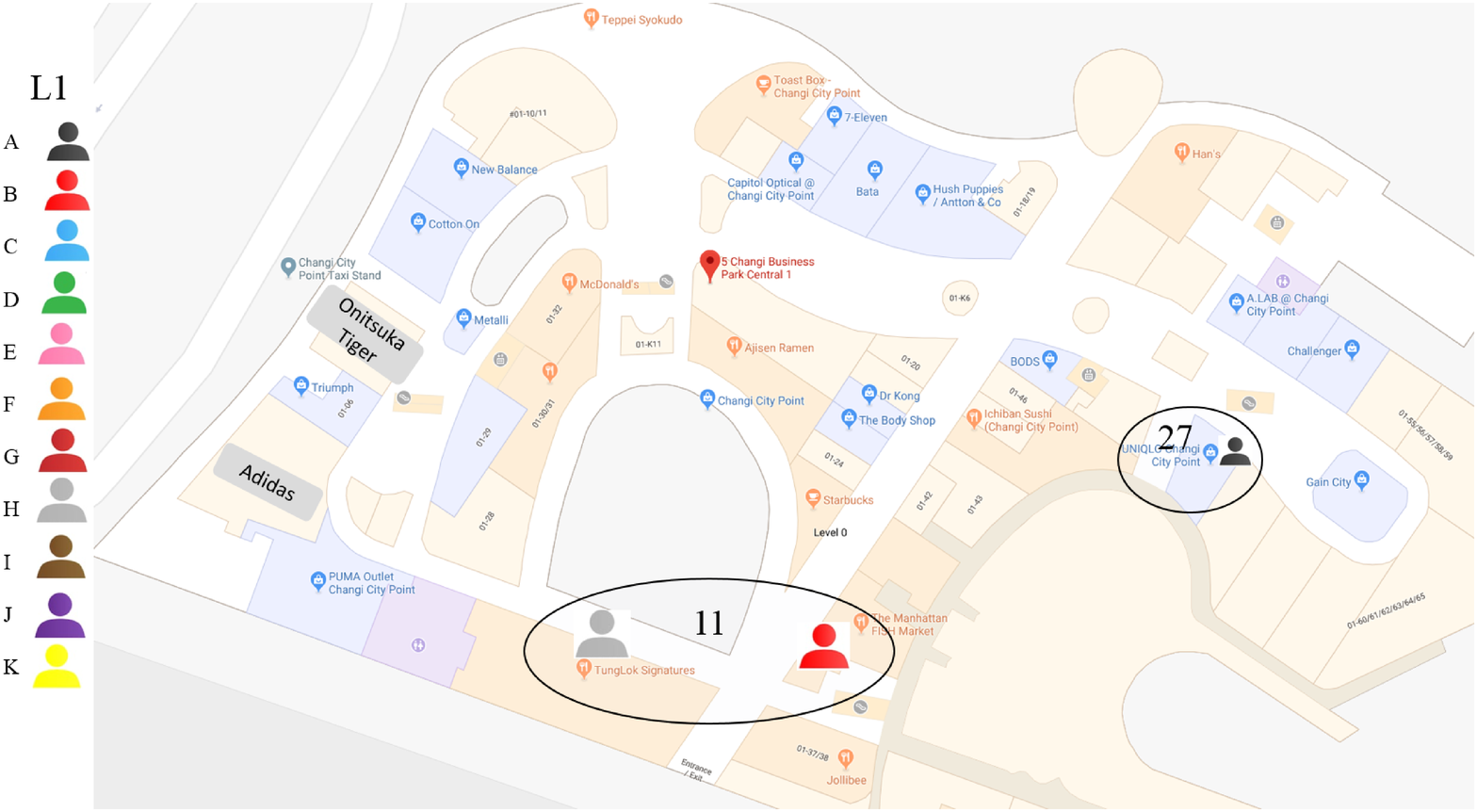}
		\caption{Level 1} \label{fig:1b}
	\end{subfigure}
	\hspace*{\fill} 
	\begin{subfigure}{0.32\textwidth}
		\includegraphics[width=\linewidth]{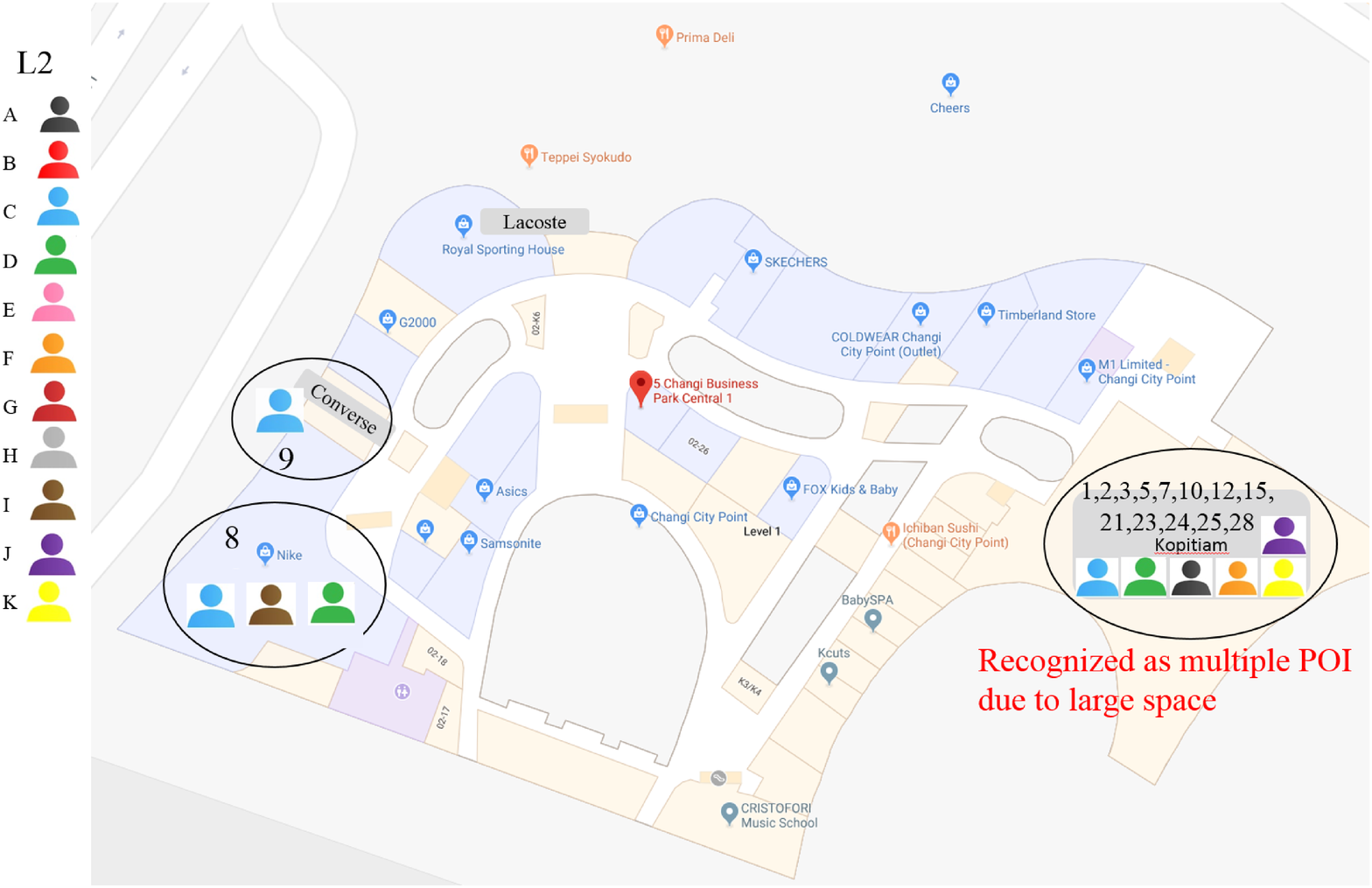}
		\caption{Level 2} \label{fig:1c}
	\end{subfigure}
	\caption{Common POI among users in three different levels of a shopping mall} \label{fig:shopping_mall}
\end{figure*}

\section{Conclusion}
\label{sec_5}

In this paper, we presented a mobile crowdsensing approach to identify indoor points of interest via Wi-Fi similarity measurements. We developed a smartphone application to collect the device's information such as GPS location and surrounding Wi-Fi access points, and use such data to identify the user's points of interest in a given indoor environment. Through an experimental study, we demonstrate that it is possible to identify the common POI among users, using community detection techniques. In future work, we plan to extend the smartphone application into iOS platform. Also, our aim is to develop a POI recommendation system and to profile users based on similar interest and analyse their indoor mobility patterns.

\section*{Acknowledgment}
This research, led together with the Housing and Development Board, is supported by the Singapore Ministry of National Development and the National Research Foundation, Prime Ministers Office under the Land and Livability National Innovation Challenge (L2 NIC) Research Programme (L2 NIC Award No. L2NICTDF1-2017-4). Any opinions, findings, and conclusions or recommendations expressed in this material are those of the author(s) and do not reflect the views of the Housing and Development Board, Singapore Ministry of National Development and National Research Foundation, Prime Ministers Office, Singapore.

\bibliographystyle{IEEEtran}

\bibliography{refer}


\end{document}